# Implicit infinite lattice summations for real space ab initio correlation methods.

M. Albrecht*

*Theoretical Chemistry FB08, University of Siegen, 57068 Siegen/Germany*

## Abstract

We suggest a local wave function based *ab initio* correlation method for infinite periodic systems, which can describe both the near range as well as the long range correlation effects coherently in the same scheme. Specifically, this work introduces a formalism which allows to describe the long range polarization cloud around a quasi particle in a solid explicitly in the formalism of local wave function based *ab initio* descriptions. To this end we reformulate the infinite lattice summation underlying the quantum chemistry formula to second order in a closed analytic form employing the elliptic theta function of the third kind. All formulas and manipulations are developed explicitely in full detail and a first numeric example demonstrates the principle idea. Good results for the long range polarization effects in LiH and LiF are found in agreement with earlier estimates.

* Corresponding author: e-mail: m.albrecht@uni-siegen.de, Phone: +49 271 740 4024, Fax: +00 271 740 2805

## I. INTRODUCTION

The field of many–body theory has been developing at high speed in recent years. A significant amount of effort is directed towards affordable descriptions of electronic correlation in infinite periodic systems or large heterogeneous systems both in the ground state and in excited states. Well established schemes like the density functional theory (DFT)[1] have seen more and more refinements, for example by means of the optimized effective potential method (OEP)[2], time dependent DFT[3], screening implementations[4], the Wigner theory[5] or the reduced density matrix functional theory (RDMFT)[6]. Other correlation schemes have been derived, examples are the dynamical mean field theory (DMFT)[7] or the GW approximation[8].

Parallel to the density functional approaches wave function based quantum chemical methods have also been trimmed to be applicable to large systems and finally even to polymers and crystals. To a large extend credit is due to the local ansatz of Fulde, Stollhoff and Horsch[9,10] as well as the systematic development of an incremental scheme by Stoll[11–13]. As for ground state properties the ansatz of this local incremental scheme in combination with the coupled cluster (CC) method turned out to be valiant in applications to broad classes of polymers and semiconductors as well as ionic crystals[11–15]. Extensions of this particular approach to band structure calculations were following suit. Gräfenstein and Stoll presented an effective Hamiltonian obtained from cluster approximations for group–IV–semiconductors[16–18], which were subsequently applied successfully to polymers as well[19–21]. Further significant advances for the case of excited states have been reported for quantum monte carlo methods (QMC)[22–25], algebraic diagrammatic construction (ADC)[26], or the Green's function approach[27–33].

These wave function based methods are straightforwardly applicable to both ground state and excited state calculations alike and are amenable to systematic improvement on the numerical accuracy by their very construction.

The general bottle–neck of steep increase of numerical effort with system size, however, affects all wave function based methods alike. It is precisely this obstacle which was overcome in earlier applications by a formulation of electron correlations in local orbitals and a hierarchy of correlation contributions called the incremental scheme[11,15–18].

Recently we demonstrated that a full Green's function approach with a frequency dependent self energy can in principle also be combined with the incremental scheme. Band structure calculations were performed for LiH and LiF[29–31] and a recent application to a molecule inside a molecular junction also underlines the usefulness of such an approach[32]. The key enabling such calculations was an approach based on local orbitals and a real space formulation of the self energy. A particular decomposition of energy denominators appearing in quantum chemical expressions led to a considerable speed–up by two orders of magnitude[33].

However, one problem pertains to all these efforts, namely the presence of a long range polarization cloud around a quasi particle[34]. This effect can shift band structures by some $eV$ and has hitherto only been estimated by parametrized continuum approximations[16,17,29–31] in the frame of the aforementioned local orbital based *ab initio* schemes. While the terms giving rise to the long range polarization of the crystal due to the presence of an extra quasi particle are well known, the underlying infinite lattice summation could not be performed so far, in particular in three dimensional systems. Of course, there are other methods available to describe the long range polarization cloud, like the random phase approximation (RPA)[34,35], but they cannot be combined with the description of near range correlation effects derived from our local orbital based correlation scheme without double counting. Due to the efficiency in describing correlation effects locally, we would like to stick to the local incremental scheme, while extending it to a description of the long range polarization cloud.

In the present work we device a way to solve this problem. The application of a mathematical identity allows to decompose the two denominators in a perturbative expression so as to prepare the ground for an infinite lattice summation performed analytically. The denominator decomposition has been applied previously in a different context and a slightly different form to energy denominators[33,36]. The ensuing lattice summation bears similarities with the well established Ewald summation technique[37,38], albeit the mathematical details differ distinctly. The main purpose of this work is to reveal all formulas in detail and to sketch the general idea. Along with this we also present a first numerical application to further exemplify the procedure.

In the following section we describe the theory with the focus on the transformation of the infinite lattice sum. In Sec. III numerical tests are presented together with applications to two realistic ionic systems, LiH and LiF. Sec. IV contains our conclusions.

## II. THEORY

In earlier works we have designed a formulation of the Green's function correlation method so as to use local HF orbitals as a starting point and then assess the correlation contributions in a full *ab initio* manner. This correlation method has subsequently been combined with the so called incremental scheme which proved to boost efficiency tremendously as was demonstrated in some solid state band structure calculations[29–32].

In earlier works an effective Hamiltonian was set up in a similar way to calculate valence band structures of semiconductors and polymers[16–19,21]. Analogous schemes were developed even earlier to determine ground state correlation energies of solids as well as polymers[11–13,15,20].

Common to all these efforts is that the correlation hole around a quasi particle is described efficiently in real space by means of a local correlation scheme based on Wannier type HF orbitals as a starting point. While this turned out to be very successful, one shortcoming has never been resolved, namely the problem of the long range polarization cloud. This is a specific type of correlation contribution in infinitely extended periodic systems which is not short range. Quasi particles like extra electrons or holes cause a polarization of the crystal which significantly lowers the overall energy of the system. In band structure calculations this effect was estimated to shift the HF valence and conduction bands in the order of $eV$ into the fundamental gap. This effect has been estimated from parametrized continuum electrostatics so far[16–19,21,29,30].

In this work we demonstrate how the long range polarization cloud can be incorporated into the procedure in a fully *ab initio* way.

We first briefly repeat the Green's function correlation method and a description of the incremental scheme in Sec. II A and II B.

In the main part Sec. II C a mathematical equality is presented to overcome the numerical obstacle of the infinite lattice summation. The quantum chemical expressions are adapted towards this formula and numerical checks are performed in Sec. III together with a first application to the band structure shifts in the LiH and LiF lattices.

### A. The Green's function

The starting point of our approach are localized occupied as well as virtual HF orbitals. While we are interested in solids, the notation is kept very general for the time being. In terms of such orbitals a model space P and excitation space Q are distinguished for the example of virtual states (the case of occupied states being completely analogous) as follows: The model space P – describing the HF level – comprises of the $(N+1)$–particle HF determinants $|\eta\rangle$, while the correlation space Q contains single and double excitations $|\alpha\rangle$ on top of $|\eta\rangle$:

$$|\eta\rangle = c_n^\dagger|\Phi_{\rm HF}\rangle, \qquad |\alpha\rangle = c_r^\dagger c_a|\eta\rangle, \quad c_r^\dagger c_s^\dagger c_a c_b|\eta\rangle \tag{1}$$

$$P = \sum_\eta |\eta\rangle\langle\eta|, \qquad Q = \sum_\alpha |\alpha\rangle\langle\alpha|. \tag{2}$$

We adopt the index convention that $a, b, c, d$ and $r, s, t, u, \eta, \mu$ represent occupied and virtual HF orbitals, respectively. The idea of locality translates into a restriction of the area the orbital can be chosen from to one or more contiguous spatial parts of the system, as exploited systematically in the incremental scheme to be sketched below. It is important to note that by enlarging the size of the spatial area thus covered, this approximation can be checked in a systematic way for convergence.

With the above notation the Green's function takes the form $G_{\eta\mu}(t) = -i\langle T[c_\eta(0)c_\mu^\dagger(t)]\rangle$, where T is the time–ordering operator and the brackets denote the average over the exact ground–state. It can be obtained from Dyson's equation as:

$$G_{\eta\mu}(\omega) = [\omega\underline{1} - F - \Sigma(\omega)]^{-1}_{\eta\mu} . \tag{3}$$

Here the self energy $\Sigma_{\eta\mu}(\omega)$ which contains the correlation effects, has been introduced and $\underline{1}$ represents the unity matrix. $G^0_{\eta\mu}(\omega)$ is the HF propagator $[G^0(\omega)]^{-1}_{\eta\mu} = \omega\delta_{\eta\mu} - F_{\eta\mu}$. The correlated energies are given by the poles of the Green's function which are numerically iteratively retrieved as the zeros of the denominator in Eq. (3).

The construction of the self energy has been described repeatedly[30? ,31]. Here we just quote the result so as to establish the link to our later formula for the long range polarization cloud.

Only the retarded self energy part is discussed, the case of the advanced part being analogous.

The space of 2–particle 1–hole states (2p1h) is spanned by $|r, s, a\rangle = a_{\rm r}^\dagger a_{\rm s}^\dagger a_{\rm a}|\Phi_{\rm HF}\rangle$. The Hamiltonian is set up in this basis as: $[H^{\rm R}]_{\rm rsa,r's'a'} = \langle r, s, a|H - E_0|r', s', a'\rangle$ and is subsequently diagonalize.

Diagonalizing the matrix $H^{\rm R}$ results in the eigenvectors $S^{\rm R}$ and eigenvalues $\lambda^{\rm R}$. The retarded part of the self energy is then constructed as

$$\begin{aligned}\Sigma^{\rm R}_{\eta\mu}(\omega) &= \sum_{\rm rsa;r's'a'} \Gamma({\rm rs};\eta{\rm a})\left[\omega\underline{1} - H^{\rm R} + i\delta\underline{1}\right]^{-1}_{\rm rsa;r's'a'}\Gamma({\rm r's'};\mu{\rm a'}) \\ &= \sum_{\rm rsa;r's'a'} \Gamma({\rm rs};\eta{\rm a})\sum_{\rm q} S^{\rm R}_{rsa;q}\frac{1}{\left(\omega - \lambda^{\rm R}_{\rm q} + i\delta\right)}S^{\rm R}_{q;r's'a'}\Gamma({\rm r's'};\mu{\rm a'}).\end{aligned} \tag{4}$$

$\Gamma$ is a shorthand for $\Gamma(\mathrm{rs};\mathrm{ta}) = W_{\mathrm{rsta}} - W_{\mathrm{rsat}}$ and $W$ are the standard two–electron integrals including the spin part. If the indices only refer to pure space orbitals, we will use $V$ instead of $W$ as is done in the next section.

In an earlier work we pointed out that the ordinary Møller–Plesset perturbation theory result (PT2) is obtained by only retaining the Fock quantities: $\lambda^{\mathrm{R}}_{\mathrm{rsa}} = \epsilon_{\mathrm{r}} + \epsilon_{\mathrm{s}} - \epsilon_{\mathrm{a}}$. A comparison of algebraic expressions and diagrams then allowed to establish the relation

$$\Sigma^{(\mathrm{PT2})}_{\eta\mu}(\omega = \epsilon_{\eta}) = H^{\mathrm{eff},(\mathrm{PT2})}_{\eta\mu} - F_{\eta\mu}, \tag{5}$$

where $H^{\mathrm{eff},(\mathrm{PT2})}_{\eta\mu}$ is the second order effective Hamiltonian.

### B. The incremental scheme

The efficiency of the procedure is derived from the application of an incremental scheme. The task of correlating electrons in a large system is broken down systematically to diagonalizations in smaller subsystems.

As subset of the incremental scheme the unit cell serves as the smallest unit, giving rise to one–cell increments, two–cell increments and so forth.

An incremental description of the matrix elements $\Sigma_{\eta\mu}$ starts with a correlation calculation in which only excitations inside one unit cell, designated to be the central one with lattice vector $\mathbf{0}$, are allowed. This results in a contribution to the correlation effects which is labeled one–cell increment (cf. Eq. (6)). In a next step the calculation is repeated with excitations being allowed in a region enlarged by one additional unit cell with real space lattice vector $\mathbf{R}_1$. The result of this calculation is denoted as $\Sigma^{\mathbf{0},\mathbf{R}_1}$ and the difference with respect to the one–cell increment $\Sigma^{\mathbf{0}}$ then isolates the effect of additional excitations involving this additional unit cell and constitutes the two–cell increment as shown in Eq. (7). This procedure can be continued to more and more unit cells. In the end the summation Eq. (8) of all increments is the final approximation to the self energy. In this work we are in particular interested in the diagonal elements $\Sigma_{\eta\eta}$ of the self energy or effective Hamiltonian, since they give the overall dispersionless shifts of the HF bands, as caused partly by the long range polarization cloud.

$$\Delta\Sigma^{\mathbf{0}}_{\eta\eta} = \Sigma^{\mathbf{0}}_{\eta\eta} \tag{6}$$

$$\Delta\Sigma^{\mathbf{0},\mathbf{R}_1}_{\eta\eta} = \Sigma^{\mathbf{0},\mathbf{R}_1}_{\eta\eta} - \Delta\Sigma^{\mathbf{0}}_{\eta\eta} - \Delta\Sigma^{\mathbf{R}_1}_{\eta\eta} \tag{7}$$

$$\Sigma_{\eta\eta} \;=\; \Delta\Sigma^{\mathbf{0}}_{\eta\eta} + \sum_{i} \Delta\Sigma^{\mathbf{0},\mathbf{R}_{\rm i}}_{\eta\eta} + \sum_{i>j} \Delta\Sigma^{\mathbf{0},\mathbf{R}_{\rm i},\mathbf{R}_{\rm j}}_{\eta\eta} + \;\ldots \tag{8}$$

The main idea of the incremental series (8) is to exploit the mainly local character of correlation corrections to HF results. This feature should manifest itself in a rapid decrease of increments both with the distance between the regions involved and with their number included in the increment. This means that only a few increments need to be calculated, yet a full account of the near range correlations is achieved this way. It is crucial to emphasize that the cutoff thus introduced in the summation (8) is well controlled, since the decrease of the incremental series can be explicitly monitored.

The shortcoming of this scheme is that the long range polarization cloud cannot be assessed in this way, because it would entail an infinite lattice summation and hence infinitely many very small increments. On the other hand, a straightforward evaluation of the self energy according to the incremental scheme would include to a very vast extend excitations which do not contribute at all to the correlation effects. Only one certain class of excitations needs to be summed up to infinity as described in the next section.

### C. Derivation of the implicit infinite lattice summation formulas

The kind of excitations responsible for the long range polarization are identifiable, and most recently we have found that such contributions lead to closed forms described by elliptic theta functions. Thus the evaluation of this type of summations can be incorporated into quantum chemistry codes in a similar fashion as the Ewald summation, and would describe the effects of the infinite lattice implicitly. For simplicity we sketch the problem in one dimension for the case of polyacetylene in Fig. 1. An extra charge in the central unit cell (the one in the foreground in the figure) gives rise to local electron–hole excitations in the neighboring unit cells. These local dipole excitations add up to what is known as long range polarization cloud and should be summed up to infinity. With the local schemes developed so far this cannot be achieved and has hence been taken into account by some continuum approximation in previous applications.

In diagrammatic notation the polarization of the crystal due to some extra charge in orbital $\eta$ is described by a local excitation from orbital $a$ to $r$ as indicated in Fig. 2.

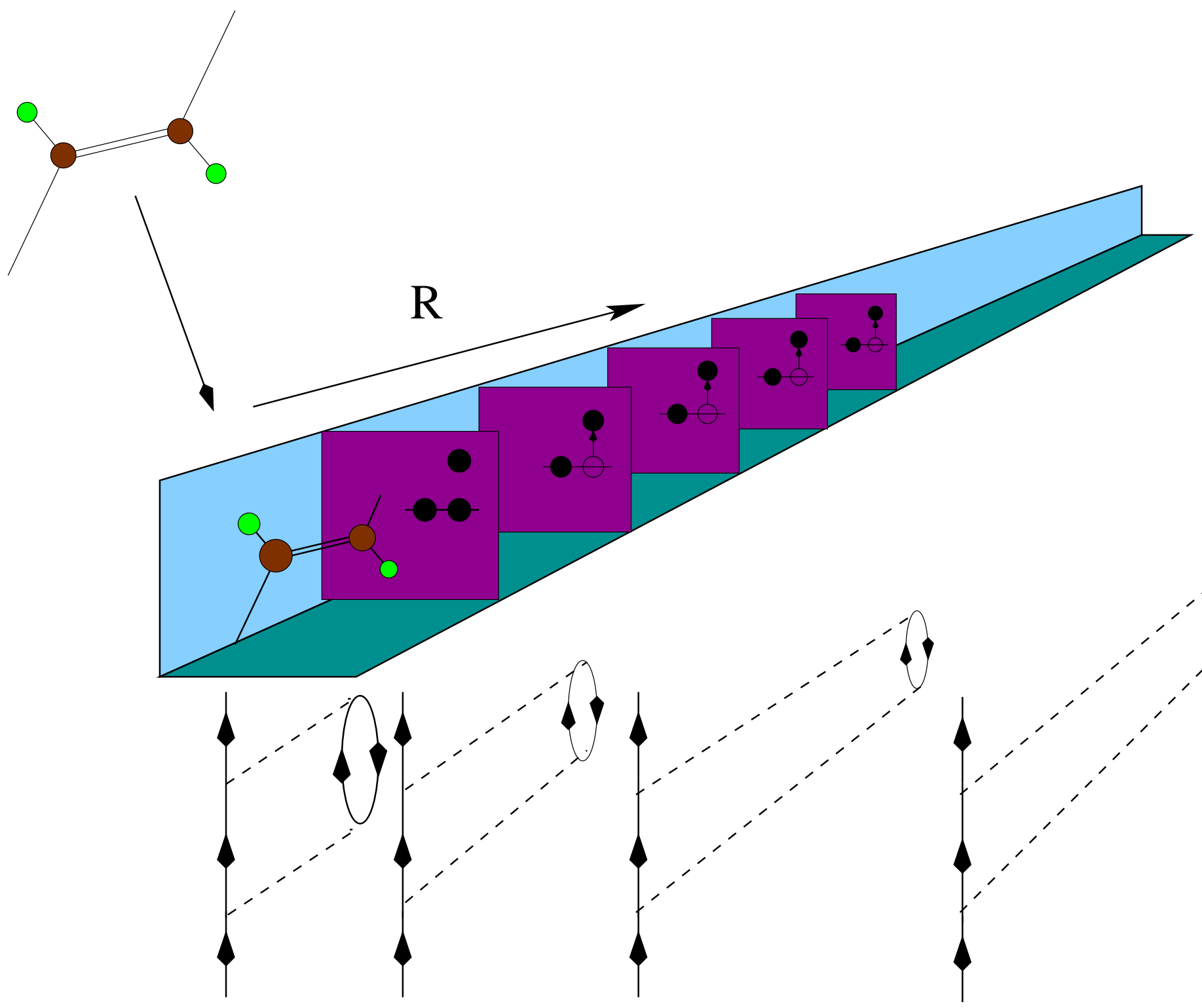

FIG. 1: *Sketch of the local dipole excitations and the underlying diagrams leading to the polarization cloud.*

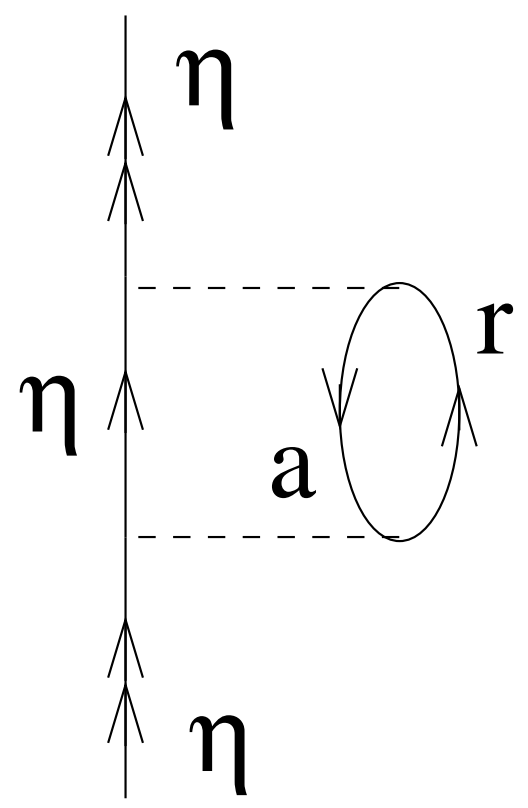

FIG. 2: *Typical representative of a diagram adding correlations due to the presence of an additional particle in orbital* $\eta$.

When switching to spin–free orbital notation and focussing on a closed shell system, the diagram translates into

$$H_{\eta\eta} = 2\sum_{\mathrm{a,r}} \frac{V_{\eta\mathrm{a}\eta\mathrm{r}} V_{\eta\mathrm{r}\eta\mathrm{a}}}{\epsilon_{\mathrm{a}} - \epsilon_{\mathrm{r}}} \tag{9}$$

for the matrix element $H_{\eta\eta}$ of an effective Hamiltonian or

$$\Sigma_{\eta\eta}(\omega) = 2\sum_{\mathrm{a,r}} \frac{V_{\eta\mathrm{a}\eta\mathrm{r}} V_{\eta\mathrm{r}\eta\mathrm{a}}}{\omega + \epsilon_{\mathrm{a}} - \epsilon_{\mathrm{r}} - \epsilon_{\eta}} \tag{10}$$

for the matrix element $\Sigma_{\eta\eta}(\omega)$ of the self energy. The two formulas are related by

$$H_{\eta\eta} = \Sigma_{\eta\eta}(\omega = \epsilon_{\eta}) \tag{11}$$

as was discussed in previous studies[30,31]. In the following it does not matter which case is pursued and we stick to the effective Hamiltonian expression (9).

We now explicitely perform the sum over all unit cells to describe local excitations from orbital $a$ to orbital $r$ in unit cell $\mathbf{R}$. Thus in the following the orbital summations only refer to the central unit cell $\mathbf{0}$, while the unit cell is indicated explicitely for each orbital index:

$$H^{\infty}_{\eta\eta} = 2\sum_{\mathrm{a,r}\in\mathbf{0}} \sum_{\mathbf{R}}^{\infty} \frac{V_{\eta(\mathbf{0})\mathrm{a}(\mathbf{R})\eta(\mathbf{0})\mathrm{r}(\mathbf{R})} V_{\eta(\mathbf{0})\mathrm{r}(\mathbf{R})\eta(\mathbf{0})\mathrm{a}(\mathbf{R})}}{\epsilon_{\mathrm{a}} - \epsilon_{\mathrm{r}}} \tag{12}$$

Due to the degeneracy with respect to the unit cell the energy denominator can be taken out of the lattice summation which then only runs over the two–electron integrals:

$$H^{\infty}_{\eta\eta} = 2\sum_{\mathrm{a,r}\in\mathbf{0}} \frac{1}{\epsilon_{\mathrm{a}} - \epsilon_{\mathrm{r}}} \sum_{\mathbf{R}}^{\infty} V_{\eta(\mathbf{0})\mathrm{a}(\mathbf{R})\eta(\mathbf{0})\mathrm{r}(\mathbf{R})} V_{\eta(\mathbf{0})\mathrm{r}(\mathbf{R})\eta(\mathbf{0})\mathrm{a}(\mathbf{R})}\,. \tag{13}$$

The index $\infty$ is a reminder that the lattice sums above are to be performed over the entire lattice and will be dropped henceforth. Inserting the representation of each local HF orbital in the basis set

$$\mid \eta > = \sum_{\alpha} C_{\alpha,\eta} \mid \alpha > \tag{14}$$

will transform the lattice sum as:

$$\sum_{\mathbf{R}} V_{\eta(\mathbf{0})\mathrm{a}(\mathbf{R})\eta(\mathbf{0})\mathrm{r}(\mathbf{R})} V_{\eta(\mathbf{0})\mathrm{r}(\mathbf{R})\eta(\mathbf{0})\mathrm{a}(\mathbf{R})} = \sum_{\alpha\beta\gamma\sigma} C_{\alpha,\eta}C_{\beta,\mathrm{a}}C_{\gamma,\eta}C_{\sigma,\mathrm{r}} \sum_{\alpha'\beta'\gamma'\sigma'} C_{\alpha',\eta}C_{\beta',\mathrm{r}}C_{\gamma',\eta}C_{\sigma',\mathrm{a}} \tag{15}$$

$$\sum_{\mathbf{R}} V_{\alpha(\mathbf{0})\beta(\mathbf{R})\gamma(\mathbf{0})\sigma(\mathbf{R})} V_{\alpha'(\mathbf{0})\beta'(\mathbf{R})\gamma'(\mathbf{0})\sigma'(\mathbf{R})}\,. \tag{16}$$

With the two–electron integrals written out the lattice sum (16) takes the form:

$$\sum_{\mathbf{R}} V_{\alpha(\mathbf{0})\beta(\mathbf{R})\gamma(\mathbf{0})\sigma(\mathbf{R})} V_{\alpha'(\mathbf{0})\beta'(\mathbf{R})\gamma'(\mathbf{0})\sigma'(\mathbf{R})} = \sum_{\mathbf{R}} \int d^3\mathbf{r}_1 d^3\mathbf{r}_2 d^3\mathbf{r}_3 d^3\mathbf{r}_4 \tag{17}$$

$$\frac{\phi_{\alpha}(\mathbf{r}_1)\phi_{\beta}(\mathbf{r}_2 - \mathbf{R})\phi_{\gamma}(\mathbf{r}_1)\phi_{\sigma}(\mathbf{r}_2 - \mathbf{R})}{\mid \mathbf{r}_1 - \mathbf{r}_2 \mid} \tag{18}$$

$$\frac{\phi_{\alpha'}(\mathbf{r}_3)\phi_{\beta'}(\mathbf{r}_4 - \mathbf{R})\phi_{\gamma'}(\mathbf{r}_3)\phi_{\sigma'}(\mathbf{r}_4 - \mathbf{R})}{\mid \mathbf{r}_3 - \mathbf{r}_4 \mid}. \tag{19}$$

*1. Denominator Decomposition*

The basic idea leading to a solution of the integral including the infinite lattice summation is to decompose the denominators into Gaussian lobes. Since the basis functions $\phi_\alpha$ are ultimately also decomposed into Gaussian lobes, the integrals can be performed analytically leaving us with a lattice sum over Gaussian type functions, which we can finally transform into special functions.

We start by representing the basis functions $\phi_\alpha$ in terms of Gaussian lobes

$$\phi_\alpha(\mathbf{r}) = \sum_{\mathrm{a}} D_{\mathrm{a},\alpha}\varphi_{\mathrm{a}}(\mathbf{r} - \mathbf{R}_{\mathrm{a},\alpha}) \tag{20}$$

and

$$\varphi_{\mathrm{a}}(\mathbf{r} - \mathbf{R}_{\mathrm{a},\alpha}) = A_\alpha e^{-g_{\mathrm{a},\alpha}(\mathbf{r}-\mathbf{R}_{\mathrm{a},\alpha})^2}, \tag{21}$$

where $A_\alpha$ is a normalization constant and the sum in (20) runs over the primitive lobes with displacements $\mathbf{R}_{\mathrm{a},\alpha}$ necessary to depict a basis function in the customary way. In the following we write for convenience:

$$g_{\mathrm{a}} := g_{\mathrm{a},\alpha} \tag{22}$$

$$\mathbf{R}_{\mathrm{a}} := \mathbf{R}_{\mathrm{a},\alpha}. \tag{23}$$

The two–electron integrals are commonly broken down to the primitive Gaussian lobe level in quantum chemistry program packages such as WANNIER[39], involving another transformation to:

$$\sum_{\mathbf{R}} V_{\alpha(\mathbf{0})\beta(\mathbf{R})\gamma(\mathbf{0})\sigma(\mathbf{R})} V_{\alpha'(\mathbf{0})\beta'(\mathbf{R})\gamma'(\mathbf{0})\sigma'(\mathbf{R})} = \sum_{\mathrm{abcd}} D_{\mathrm{a}\alpha}D_{\mathrm{b}\beta}D_{\mathrm{c}\gamma}D_{\mathrm{d}\sigma} \sum_{\mathrm{a'b'c'd'}} D_{\mathrm{a'}\alpha'}D_{\mathrm{b'}\beta'}D_{\mathrm{c'}\gamma'}D_{\mathrm{d'}\sigma'} \tag{24}$$

$$\sum_{\mathbf{R}} V_{\mathrm{a}(\mathbf{0})\mathrm{b}(\mathbf{R})\mathrm{c}(\mathbf{0})\mathrm{d}(\mathbf{R})} V_{\mathrm{a'}(\mathbf{0})\mathrm{b'}(\mathbf{R})\mathrm{c'}(\mathbf{0})\mathrm{d'}(\mathbf{R})}. \tag{25}$$

This leaves us with the task of evaluating and summing the product of two two–electron integrals. This part of Eq. (25) takes the form:

$$\sum_{\mathbf{R}} V_{\mathrm{abcd}}V_{\mathrm{a'b'c'd'}} = \sum_{\mathbf{R}} \int d^3\mathbf{r}_1 d^3\mathbf{r}_2 d^3\mathbf{r}_3 d^3\mathbf{r}_4 \tag{26}$$

$$\frac{\varphi_{\mathrm{a}}(\mathbf{r}_1)\varphi_{\mathrm{b}}(\mathbf{r}_2 - \mathbf{R})\varphi_{\mathrm{c}}(\mathbf{r}_1)\varphi_{\mathrm{d}}(\mathbf{r}_2 - \mathbf{R})}{\mid \mathbf{r}_1 - \mathbf{r}_2 \mid} \tag{27}$$

$$\frac{\varphi_{\mathrm{a'}}(\mathbf{r}_3)\varphi_{\mathrm{b'}}(\mathbf{r}_4 - \mathbf{R})\varphi_{\mathrm{c'}}(\mathbf{r}_3)\varphi_{\mathrm{d'}}(\mathbf{r}_4 - \mathbf{R})}{\mid \mathbf{r}_3 - \mathbf{r}_4 \mid}. \tag{28}$$

We now turn to the decomposition of the denominators appearing in Eq. (27,28) according to a formula analyzed by Hackbusch and Khoromskij[40]:

$$\frac{1}{\rho} = h \sum_{m=-l}^{l} f(mh) e^{-\rho^2 g(mh)}, \tag{29}$$

where $h$ is a suitable integration width and $f, g$ are suitable functions so as to make the series rapidly converge to machine precision. In fact, we found that about 30 terms ($l = 32$ in Eq. (29)) are enough to ensure satisfactory precision in all cases. The functions $f, g$ can be taken from ref.[40], where extensive studies concerning this decomposition are available, and have the explicit forms:

$$f(\rho) = \frac{2}{\sqrt{\pi}} \frac{cosh(\rho)}{1 + e^{-sinh(\rho)}} \tag{30}$$

$$g(\rho) := log^2(1 + e^{sinh(\rho)}). \tag{31}$$

A similar decomposition has been studied extensively by us in a previous work[33]. Further studies to the numerical accurateness of truncating expansion (29) will be presented in Sec. III.

With the abbreviations

$$f(mh) =: f_{\mathrm{m}} \;\; ; \;\; g(mh) =: g_{\mathrm{m}} \tag{32}$$

the integral over one typical expression originating from (27) together with the expansion (20) takes the form of an integral over five Gaussian lobes:

$$\begin{aligned} V_{\mathrm{abcd}} = {} & A_{\mathrm{a}} A_{\mathrm{b}} A_{\mathrm{c}} A_{\mathrm{d}} h \sum_{\mathrm{m}} f_{\mathrm{m}} \int d^3\mathbf{r}_1 d^3\mathbf{r}_2 \\ & e^{-g_{\mathrm{a}}(\mathbf{r}_1-\mathbf{R}_{\mathrm{a}})^2} e^{-g_{\mathrm{c}}(\mathbf{r}_1-\mathbf{R}_{\mathrm{c}})^2} e^{-g_{\mathrm{b}}(\mathbf{r}_2-[\mathbf{R}_{\mathrm{b}}+\mathbf{R}])^2} e^{-g_{\mathrm{d}}(\mathbf{r}_2-[\mathbf{R}_{\mathrm{d}}+\mathbf{R}])^2} e^{-g_{\mathrm{m}}(\mathbf{r}_2-\mathbf{r}_1)^2}, \end{aligned} \tag{33}$$

where $\mathbf{R}$ is the summation index of the original infinite lattice summation. This double integral over five Gaussians is straightforward to solve. Applying formula (62) from appendix A yields a single Gaussian with respect to $\mathbf{R}$ and some lengthy prefactors as:

$$V_{\mathrm{abcd}} = A_{\mathrm{a}} A_{\mathrm{b}} A_{\mathrm{c}} A_{\mathrm{d}} h \sum_{\mathrm{m}} f_{\mathrm{m}} \frac{\pi^3}{g_0^{\frac{3}{2}}} e^{-\frac{1}{g_0}\left[D_0 + D_{\mathbf{R}} - 2C_{\mathrm{a,c}} - 2Q_{\mathrm{b,d}} - 2L_{\mathrm{a,b}} - 2L_{\mathrm{a,d}} - 2L_{\mathrm{c,b}} - 2L_{\mathrm{c,d}}\right]}, \tag{34}$$

where we used the abbreviations

$$\begin{aligned} g_0 &= (g_{\mathrm{a}} + g_{\mathrm{c}})(g_{\mathrm{b}} + g_{\mathrm{d}}) + g_{\mathrm{m}}(g_{\mathrm{a}} + g_{\mathrm{c}} + g_{\mathrm{b}} + g_{\mathrm{d}}) \\ D_0 &= D^{\mathrm{c;bd}} g_{\mathrm{a}} \mathbf{R}_{\mathrm{a}}{}^2 + D^{\mathrm{a;bd}} g_{\mathrm{c}} \mathbf{R}_{\mathrm{c}}{}^2 \\ D_{\mathbf{R}} &= D^{\mathrm{d;ac}} g_{\mathrm{b}} (\mathbf{R}_{\mathrm{b}} + \mathbf{R})^2 + D^{\mathrm{b;ac}} g_{\mathrm{d}} (\mathbf{R}_{\mathrm{d}} + \mathbf{R})^2 \end{aligned} \tag{35}$$

$$\begin{aligned}
C_{\mathrm{a,c}} &= g_{\mathrm{a}} g_{\mathrm{c}} (g_{\mathrm{b}} + g_{\mathrm{d}} + g_{\mathrm{m}}) \mathbf{R}_{\mathrm{c}} \mathbf{R}_{\mathrm{a}} \\
Q_{\mathrm{b,d}} &= g_{\mathrm{b}} g_{\mathrm{d}} (g_{\mathrm{a}} + g_{\mathrm{c}} + g_{\mathrm{m}}) (\mathbf{R}_{\mathrm{b}} + \mathbf{R})(\mathbf{R}_{\mathrm{d}} + \mathbf{R}) \\
L_{\mathrm{a,b}} &= g_{\mathrm{m}} g_{\mathrm{a}} g_{\mathrm{b}} \mathbf{R}_{\mathrm{a}} (\mathbf{R}_{\mathrm{b}} + \mathbf{R}) \\
L_{\mathrm{a,d}} &= g_{\mathrm{m}} g_{\mathrm{a}} g_{\mathrm{d}} \mathbf{R}_{\mathrm{a}} (\mathbf{R}_{\mathrm{d}} + \mathbf{R}) \\
L_{\mathrm{c,b}} &= g_{\mathrm{m}} g_{\mathrm{c}} g_{\mathrm{b}} \mathbf{R}_{\mathrm{c}} (\mathbf{R}_{\mathrm{b}} + \mathbf{R}) \\
L_{\mathrm{c,d}} &= g_{\mathrm{m}} g_{\mathrm{c}} g_{\mathrm{d}} \mathbf{R}_{\mathrm{c}} (\mathbf{R}_{\mathrm{d}} + \mathbf{R}).
\end{aligned}$$

Furthermore the shorthand

$$D^{\mathrm{i;jk}} = g_{\mathrm{i}}(g_{\mathrm{j}} + g_{\mathrm{k}}) + g_{\mathrm{m}}(g_{\mathrm{i}} + g_{\mathrm{j}} + g_{\mathrm{k}}); \qquad i, j, k \in \{a, b, c, d\} \tag{36}$$

was used.

Checking the symmetry helps to make this formula plausible. It should be noted that $g_{\mathrm{a}}$, $g_{\mathrm{b}}$, $g_{\mathrm{c}}$ and $g_{\mathrm{d}}$ refer to Gaussian lobe exponents of the basis set, while $g_{\mathrm{m}}$ originates from the denominator decomposition (29, 32).

A formula completely analogous to (33) can be imagined for the term with primed indices $V_{a'b'c'd'}$ so that the lattice sum

$$\sum_{\mathbf{R}} V_{abcd} V_{a'b'c'd'} \tag{37}$$

from Eq. (26) can finally be written as a sum over Gaussians of $\mathbf{R}$:

$$\sum_{\mathbf{R}} V_{abcd} V_{a'b'c'd'} = \sum_{\mathrm{m,m'}} h^2 f_{\mathrm{m}} f_{\mathrm{m'}} J \sum_{\mathbf{R}} e^{-\mathrm{G}(\mathbf{R}+\mathbf{\Gamma})^2}. \tag{38}$$

The parameters J, G and $\Gamma$ are given by the quantities defined in (34,36) used to describe $V_{abcd}$ together with their primed counterparts to be imagined for $V_{a'b'c'd'}$. The result is stated in Eq. (69–72) in appendix B.

### 2. *Lattice summation*

We now turn to the lattice summation contained in Eq. (38). For convenience we first confine the discussion to the one–dimensional case along the $x$–axis, so that $\mathbf{R} = n_{\mathrm{x}} a$ and $\mathbf{\Gamma} = \Gamma_{\mathrm{x}}$ with $a$ taken to be the lattice constant. Eq. (38) then takes the form

$$V(\Gamma_{\mathrm{x}}) = \sum_{\mathrm{m}} h f_{\mathrm{m}} J \sum_{n_{\mathrm{x}}=-\infty}^{\infty} e^{-G(n_{\mathrm{x}} a + \Gamma_{\mathrm{x}})^2}. \tag{39}$$

To the best of our knowledge this type of summation cannot be identified with a special function. The problem is the shift $\Gamma_x$ in the exponent, which is added to the summation index $n_x a$ before the square is performed. But with one special twist the summation can be transformed as follows.

It is important to note that the function $V(\Gamma_x)$ is translationally invariant, i. e. $V(\Gamma_x + n_x a) = V(\Gamma_x)$. Consequently this function can be Fourier–transformed resulting in:

$$V(\Gamma_x) = \sum_{m} h f_m \frac{J}{a}\sqrt{\frac{\pi}{G}} \sum_{k_x=-\infty}^{\infty} e^{-\frac{\pi^2}{a^2 G}k_x{}^2} e^{-\frac{2\pi i}{a}k_x \Gamma_x}. \tag{40}$$

The latter step is decisive, since now the summation is in a form matching a special function. In fact the elliptic theta function of the third kind $\vartheta_3$ is defined as[41]:

$$\vartheta_3(u;q) := \sum_{n=-\infty}^{\infty} q^{n^2} e^{2inu}. \tag{41}$$

The sum in reciprocal space can thus be expressed as

$$\sum_{k_x=-\infty}^{\infty} e^{-\frac{\pi^2}{a^2 G}k_x{}^2} e^{-\frac{2\pi i}{a}k_x \Gamma_x} = \vartheta_3\left(\frac{\pi\Gamma_x}{a}; e^{-\frac{\pi^2}{a^2 G}}\right), \tag{42}$$

and we finally resolve the lattice summation (38) as:

$$\sum_{\mathbf{R}} V_{abcd} V_{a'b'c'd'} = \sum_{m,m'} h^2 f_m f_{m'} \frac{J}{a}\sqrt{\frac{\pi}{G}} \vartheta_3\left(\frac{\pi\Gamma_x}{a}; e^{-\frac{\pi^2}{a^2 G}}\right). \tag{43}$$

We also state the straightforward extension to simple cubic three–dimensional lattices for later use in an obvious notation:

$$\sum_{\mathbf{R}} V_{abcd} V_{a'b'c'd'} = \sum_{m,m'} h^2 f_m f_{m'} \frac{J}{\Omega}\sqrt{\frac{\pi}{G}} \vartheta_3\left(\frac{\pi\Gamma_x}{a_x}; e^{-\frac{\pi^2}{a_x^2 G}}\right)\vartheta_3\left(\frac{\pi\Gamma_y}{a_y}; e^{-\frac{\pi^2}{a_y^2 G}}\right)\vartheta_3\left(\frac{\pi\Gamma_z}{a_z}; e^{-\frac{\pi^2}{a_z^2 G}}\right), \tag{44}$$

where $\Omega$ is the volume of the unit cell in real space. Again the parameters J, G and $\Gamma = (\Gamma_x, \Gamma_y, \Gamma_z)$ are determined by the parameters of the underlying Gaussian lobes indexed with (a,b,c,d,a',b',c',d') in the way put forth in Eq. (69–72) in appendix B. The progress gained by this result with respect to the original problem is that an infinite summation in three dimensions has been changed into a brief summation of only a few terms indexed by $m, m'$ from the denominator decomposition. In this way an infinite expression has been transformed into a feasible finite one.

For the sake of comparison with earlier estimates of the shift caused by the long range polarization cloud the near range effects already accounted for in the respective *ab initio* treatment have to be subtracted from the summation over the lattice appearing in Eq. (44) and Eq. (12). Specifically, if correlation effects had been accumulated up to a certain range $R_{\mathrm{c}}$ including unit cells in a volume $\Upsilon$, the respective vectors have to be excluded in the infinite lattice sum. Thus the remaining band shift $\gamma^{\eta}_{\mathrm{pol}}$ for band index $\eta$ due to the long range polarization cloud is obtained to second order from subtracting the respective part from Eq. (12):

$$\begin{aligned}\gamma^{\eta}_{\mathrm{pol}} &:= H^{\infty}_{\eta\eta} - H^{\Upsilon}_{\eta\eta} \qquad (45)\\ &= 2\sum_{\mathrm{a,r}\in\mathbf{0}}\sum_{\mathbf{R}}^{\infty}\frac{V_{\eta(\mathbf{0})\mathrm{a}(\mathbf{R})\eta(\mathbf{0})\mathrm{r}(\mathbf{R})}V_{\eta(\mathbf{0})\mathrm{r}(\mathbf{R})\eta(\mathbf{0})\mathrm{a}(\mathbf{R})}}{\epsilon_{\mathrm{a}}-\epsilon_{\mathrm{r}}} - 2\sum_{\mathrm{a,r}\in\mathbf{0}}\sum_{\mathbf{R}}^{\Upsilon}\frac{V_{\eta(\mathbf{0})\mathrm{a}(\mathbf{R})\eta(\mathbf{0})\mathrm{r}(\mathbf{R})}V_{\eta(\mathbf{0})\mathrm{r}(\mathbf{R})\eta(\mathbf{0})\mathrm{a}(\mathbf{R})}}{\epsilon_{\mathrm{a}}-\epsilon_{\mathrm{r}}},\end{aligned}$$

where in the first sum of the last line the lattice summation runs over the infinite lattice and is performed as described in this work, and the second sum runs over the near range treated by the incremental scheme as in previous works. The latter sum is evaluated explicitely.

As a result the overall shift $\gamma^{\eta}_{\mathrm{pol}}$ of band $\eta$ due to the long range polarization effects is isolated and can be compared with previous estimates as is done in the next section.

## III. RESULTS AND DISCUSSION

While the main purpose of this work is to put forth the mathematical idea with all formulas worked out, we would also like to further exemplify our ansatz with a first numerical application. Full scale quantum chemistry routines, however, have not been implemented for this work so far due to the tremendous programming efforts typical of *ab initio* codes. The approximations presented here, however, yield very encouraging results and focus on demonstrating the feasibility of the suggested implicit lattice summation in Sec. III B. First some numerical checks are presented in Sec. III A.

### A. Numerical checks

In this section we check the denominator decomposition (29) by comparison with an analytic result. Specifically, in one dimension, the expression

$$\sum_{\mathbf{R}\neq\mathbf{0}}\frac{1}{|\mathbf{r}-\mathbf{R}|}\frac{1}{|\mathbf{r}'-\mathbf{R}|} \qquad (46)$$

converges for $\mathbf{r} = \mathbf{r}' = \mathbf{0}$. Applying our method of denominator decomposition, subsequent transformation to reciprocal space and identification with the elliptic theta function of the third

kind yields:

$$\sum_{\mathbf{R}\neq\mathbf{0}} \frac{1}{|\mathbf{r}-\mathbf{R}|}\frac{1}{|\mathbf{r}'-\mathbf{R}|} \to \sum_{\mathrm{n}\neq 0}\frac{1}{n^2} \tag{47}$$

$$\approx h^2 \sum_{\mathrm{m,m'}=-\mathrm{l}}^{\mathrm{l}} f_{\mathrm{m}} f_{\mathrm{m'}} \sqrt{\frac{\pi}{g_{\mathrm{m}}+g_{\mathrm{m'}}}}\,\vartheta_3(0; e^{-\frac{\pi^2}{g_{\mathrm{m}}+g_{\mathrm{m'}}}}).$$

On the other hand we know from Riemann's Zeta function[42] $\zeta(z)$ that

$$\sum_{\mathrm{n}\neq 0}\frac{1}{n^2} = 2\zeta(2) = \frac{\pi^2}{3} \approx 3.28987. \tag{48}$$

Tab. I summarizes the approximate results for various summation limits $l$ in Eq. (47). For as small a value as $l = 16$ the error is already less than 0.2%. In the following calculations all results were obtained with $l = 32$.

TABLE I: *Numerical approximations of Riemann's Zeta function* $2\zeta(2) \approx 3.28987$ *for various summation limits l.*

| summation limit l | 4 | 8 | 16 | 32 | 64 |
|---|---|---|---|---|---|
| approximation | 2.07754 | 3.05315 | 3.28349 | 3.28986 | 3.28987 |

Instead of an infinite lattice summation one is thus left with a double sum over only a very finite number of terms in Eq. (44). This is the merit of the optimized denominator decomposition described in Eq. (29–31). Symmetry considerations and numerical cutoff criteria will eventually reduce the summation in Eq. (44) further.

### B. Results for the LiH Crystal

In earlier studies we have presented *ab initio* studies on the band structure of LiH both with the effective Hamiltonian method[31] as well as with the Green's function approach[29]. By virtue of the incremental scheme the local correlation effects up to third nearest neighbor unit cells were taken into account in a valence double $\zeta$ (VDZ) basis set. Thus the near range volume designated $\Upsilon$ in Eq. (45) contains 43 unit cells.

The effect of the long range polarization cloud on the band structure, on the other hand, was not taken into account on an *ab initio* level. Instead, an electrostatic continuum approximation was used to estimate $\gamma^{\eta}_{\rm pol}$ defined in Eq. (45) and an additional shift of the lowest lying conduction band by 1.0 $eV$ – irrespective of the basis set, i. e. assuming a complete basis – was found.

In the following we will apply the mechanism developed here to obtain this shift from first principles. LiH is chosen for various reasons. First, it is both experimentally and theoretically well studied. Secondly, the HF gap of 13 $eV$ is severely reduced by correlation effects to 5 $eV$, thus placing LiH between the two prototype semiconductors diamond and silicon. Last of all, it is well suited for pioneering *ab initio* studies due to the small size of the unit cell with 4 electrons. We take the lattice constant $a$ from experiment to be 4.06 Å[43].

In order to apply our scheme of infinite lattice summation, the crystal is alternatively described as simple cubic with 8 atoms at each Bravais lattice sight, rather than a fcc lattice. The unit cell thus consists of the following ions at the respective positions in units of $a/2$:

$$\begin{array}{llll} \mathbf{Li}_1 = (0,0,0) & \mathbf{Li}_2 = (1,0,1) & \mathbf{Li}_3 = (1,1,0) & \mathbf{Li}_4 = (0,1,1) \\ \mathbf{H}_1 = (0,0,1) & \mathbf{H}_2 = (1,0,0) & \mathbf{H}_3 = (1,1,1) & \mathbf{H}_4 = (0,1,0). \end{array} \quad (49)$$

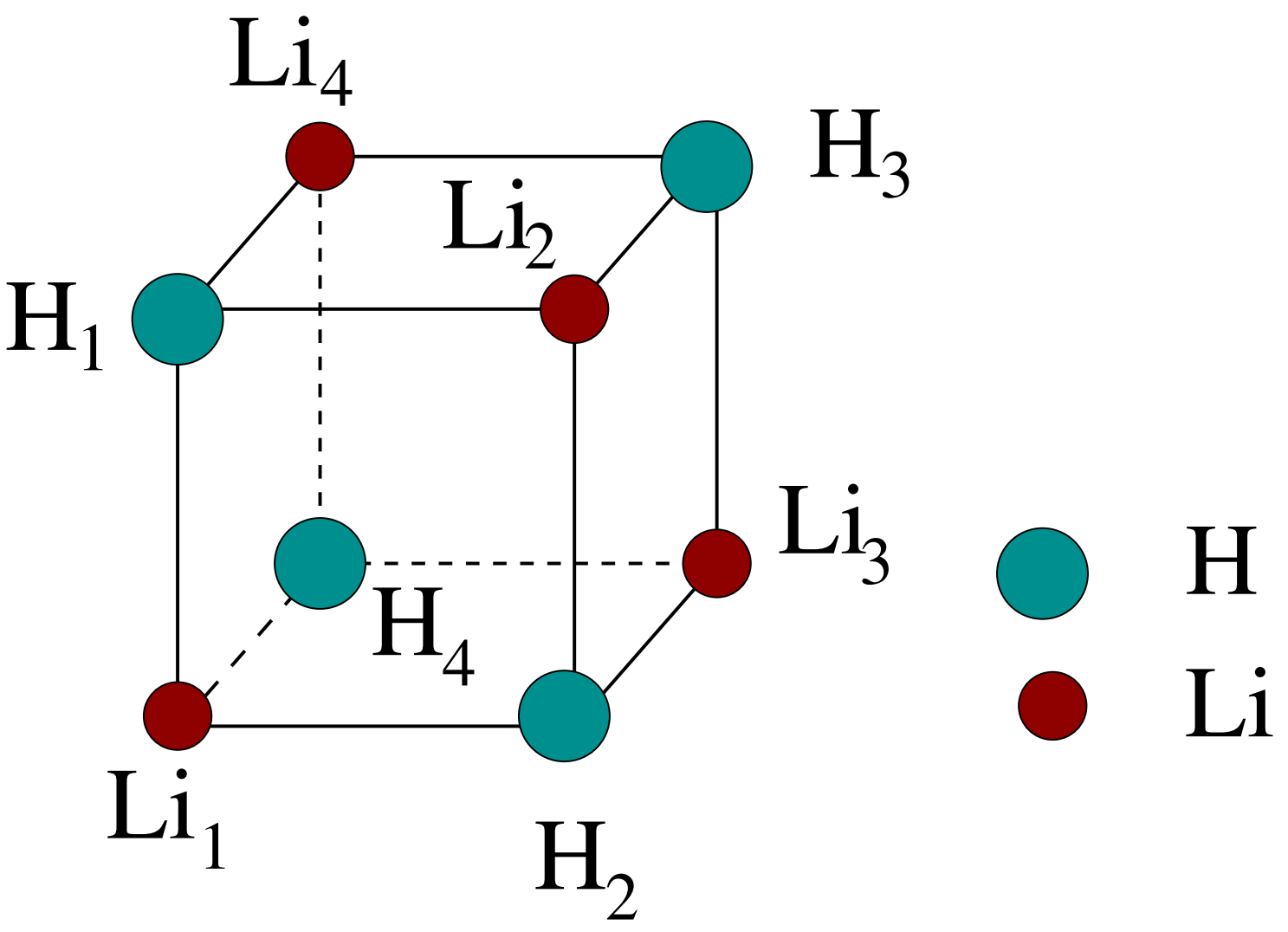

FIG. 3: Sketch of the primitive unit cell of the LiH crystal. The ions are numbered according to the vectors in Eq. (49).

The situation is sketched in Fig. 3. Starting from an ionic picture, the 1s orbitals both at the Li and at the H are doubly occupied. The lowest lying conduction band at the fundamental gap then arises from the 2s orbital of the Li cation. In the following we identify the index $\eta$ from the formulas above with this band. The presence of this extra electron at the site of $Li_1$ causes a polarization of the crystal which is basically due to polarizations of the extended, doubly occupied 1s cloud of the H anions. Thus the state $\eta$ in the notation of Fig. 2 is identified with the Li 2s orbital at the origin, and the pair $a, r$ describes a local excitation at a hydrogen site from the H 1s orbital ($= a$) to higher lying polarization functions like 2p, 3p, 3d, *etc.* for orbital $r$. In order to arrive at first numerical results so as to demonstrate the proposed method, the set–up of a full quantum chemistry *ab initio* code is avoided here. Rather, we approximate the HF orbitals with a simple representation in terms of Gaussian lobes. Apart from the aspect of numerical savings, this might also be feasible for future studies on the ground that at long distances such an approximation might be justified. At short ranges, the contributions had been assessed with the full *ab initio* machinery, and the simple Gaussian lobe approximation in the short range neighborhood is subtracted from the infinite lattice summation by hand anyway.

To proceed, we performed a solid state HF calculation with a valence quadruple–zeta (VQZ) basis using the aforementioned solid state program package WANNIER[39]. This gives us the energies used in the denominator of Eq. (9). Rather then using the HF Wannier–type orbitals themselves in the transformations (15,24), we use simple lobes to mimic these orbitals as discussed above, thus saving the expensive transformations. We again stress that our goal here is to obtain first numerical results for the infinite lattice summation rather than presenting full quantum chemical *ab initio* calculations.

With this in mind we evaluate the polarization diagram of Fig. 2 with the orbital identifications as discussed above. The numerator in the second order perturbation expression is obtained from Eq. (44) with the understanding, that up to third nearest neighbors the real space contributions have been evaluated and subtracted from hand according to Eq. (45). Excitations from the H 1s to higher shell s orbitals are not considered, since they would not polarize the hydrogen ion. We consider excitations into 2p, 3p, 4p as well as 3d and 4d. The results are stated for each shell in terms of a shift $\gamma_{\rm pol}$ of the lowest conduction band (originating from the Li 2s orbital) in Tab. II.

TABLE II: *Shift $\gamma_{\rm pol}$ of the lowest conduction band of LiH towards the gap due to the long range polarization cloud in eV.*

| | $2^{nd} shell$ | $3^{rd} shell$ | $4^{th} shell$ | $\Sigma$ | $C$ |
|---|---|---|---|---|---|
| $\gamma_{\rm pol}$ : | 0.17 | 0.40 | 0.19 | 0.76 | 1.02 |

The last column repeats the result of the continuum approximation[29,31] of 1.02 $eV$. The column labeled $\Sigma$ states the result of this work. We obtain a shift due to the long range polarization cloud of 0.76 $eV$ or 3/4 of the continuum estimation. Given the approximate description of the wavefunctions and the fact that our formulas are perturbative results to second order, we believe this is in very good agreement. The assignment of the various contributions to different shells of course should not be overinterpreted in light of the approximations. Still we interpret the fact that the third shell seems to give the largest contribution as an indication that the d orbitals are not negligible in describing the polarization of the hydrogen anion.

### C. Results for the LiF Crystal

LiF has the same geometric structure as LiH, with the experimental lattice constant to be 3.990 Å[44]. In earlier works we presented extensive studies of this system, including band structure calculations[30,31]. We found very good agreement with respect to experiment as well as other calculations[45–47]. However, again the effects of the long range polarization cloud were replaced by a continuum approximation and found to cause a shift of the conduction bands of 0.225 $eV$[30]. We employ the same valence double $\zeta$ plus polarization function basis set as in our earlier calculations[30,31], including a d–function at the F and a p–function at the Li site. The orbital energies of the localized HF solutions are obtained with WANNIER, the orbitals themselves are again approximated in a simple way by their most important Gaussian lobe basis function.

As in the previous section we consider the case of conduction bands. One additional electron is put into the 2s–like orbital at the first Li cation of the central unit cell, leading to a polarization of the F anions. This polarization is described in the present basis set by an excitation of the occupied 2s– and 2p–like orbitals into the virtual orbitals of the third shell.

TABLE III: *Shift $\gamma_{\text{pol}}$ of the lowest conduction band of LiH towards the gap due to the long range polarization cloud in eV.*

| | excitations from 2s | excitations from 2p | $\Sigma$ | $C$ |
|---|---|---|---|---|
| $\gamma_{\text{pol}}$ : | 0.123 | 0.095 | 0.218 | 0.225 |

The results are summarized in Tab. III. This time, excitations into the 3s–type orbitals at the F also have to be taken into account, because an excitation from 2p to 3s leads to a polarization of the charge distribution. Overall excitations out of the 2s occupied orbitals contribute $0.123$ $eV$, excitations from the 2p–type orbitals contribute 0.095 $eV$. In view of our approximation of the full HF wavefunctions care has to be taken so as not to overinterpret the details of these findings. Nonetheless we would like to comment that the somewhat larger contribution from the excitations out of the 2s–band with respect to 2p–excitations is peculiar at first sight. Yet it deserves mentioning that the 2p–bands of the F anion displayed a counterintuitive behaviour with respect to correlation effects in earlier calcualations as well[30,31,45,46,48], where correlation effects were found to result in a broadening of these bands rather then a flattening.

Overall, a shift of $0.218$ $eV$ is found, while the continuum approximation yielded a somewhat larger guess of $0.225$ $eV$. So our PT2 result comes close to the estimated value. With our calculation we thus also recover the fact that the polarization effect is much less pronounced in LiF as it is in LiH. The reason is the unusually large polarizability of the H anion.

## IV. CONCLUSIONS

In conclusion we have presented a formalism which allows to include a description of the long range polarization cloud into an overall local wave function based *ab initio* correlation scheme for infinitely extended periodic systems. It should be emphasized that with this progress all parts of the correlation effects can be assessed within the same local orbital scheme, so that all contributions are cleanly evaluated on the same footing with double counting avoided. While the previously well established local orbital based incremental scheme was very efficient in describing the near range polarization cloud, it can now be extended to also account for the long range polarization effects for the first time.

The feasibility of the infinite lattice summation formulas were demonstrated in a first numerical application for the LiH and LiF crystals and good agreement with previous estimations was found.

We believe that this constitutes a major step ahead in the design of a general local orbital based *ab initio* scheme for solids.

## V. ACKNOWLEDGMENTS

The author is grateful for an inspiring discussion with Boris N. Khoromskij from the Max–Planck–Institute for Mathematics in the Sciences (Leipzig/Germany). Support from the German

Research Foundation (Deutsche Forschungsgemeinschaft DFG) in the frame of the priority program SPP 1145 is also appreciated.

## VI. APPENDIX A: DERIVATION OF EQ. (34,35)

We now prove that

$$V_{\mathrm{abcd}} = A_{\mathrm{a}}A_{\mathrm{b}}A_{\mathrm{c}}A_{\mathrm{d}}h\sum_{\mathrm{m}} f_{\mathrm{m}}\frac{\pi^3}{g_0^{\frac{3}{2}}}e^{\left(-\frac{1}{g_0}\varepsilon\right)}, \tag{50}$$

where the exponent $\varepsilon$ takes the form

$$\begin{aligned}\varepsilon = & \\ & D^{\mathrm{c;bd}}g_{\mathrm{a}}{\mathbf{R}_{\mathrm{a}}}^2 + D^{\mathrm{a;bd}}g_{\mathrm{c}}{\mathbf{R}_{\mathrm{c}}}^2 + D^{\mathrm{d;ac}}g_{\mathrm{b}}{\mathbf{R}_{\mathrm{b}}}^2 + D^{\mathrm{b;ac}}g_{\mathrm{d}}{\mathbf{R}_{\mathrm{d}}}^2 \\ & -2g_{\mathrm{a}}g_{\mathrm{c}}(g_{\mathrm{b}} + g_{\mathrm{d}} + g_{\mathrm{m}})\mathbf{R}_{\mathrm{c}}\mathbf{R}_{\mathrm{a}} \\ & -2g_{\mathrm{b}}g_{\mathrm{d}}(g_{\mathrm{a}} + g_{\mathrm{c}} + g_{\mathrm{m}})\mathbf{R}_{\mathrm{b}}\mathbf{R}_{\mathrm{d}} \\ & -2g_{\mathrm{m}}\left[B_{\mathrm{ab}} + B_{\mathrm{ad}} + B_{\mathrm{cb}} + B_{\mathrm{cd}}\right].\end{aligned} \tag{51}$$

We have again used the shorthand (36) for the $D^{\mathrm{i;jk}}$ and

$$B_{\mathrm{ij}} = g_{\mathrm{i}}g_{\mathrm{j}}\mathbf{R}_{\mathrm{i}}\mathbf{R}_{\mathrm{j}}; \qquad i,j \in \{a,b,c,d\}. \tag{52}$$

First the integral in Eq. (33) is solved with respect to $\mathbf{r}_2$. Omitting $\mathbf{R}$ by defining

$$\mathbf{r}_{\mathrm{b}} := \mathbf{R}_{\mathrm{b}} + \mathbf{R}, \quad \mathbf{r}_{\mathrm{d}} := \mathbf{R}_{\mathrm{d}} + \mathbf{R} \tag{53}$$

(for convenience of writing), this part of the integral takes the form:

$$A_{\mathrm{b}}A_{\mathrm{d}}\int d^3\mathbf{r}_2 e^{-g_{\mathrm{b}}(\mathbf{r}_2-\mathbf{r}_{\mathrm{b}})^2}e^{-g_{\mathrm{d}}(\mathbf{r}_2-\mathbf{r}_{\mathrm{d}})^2}e^{-g_{\mathrm{m}}(\mathbf{r}_2-\mathbf{r}_1)^2}. \tag{54}$$

Rearranging the exponent yields

$$\begin{aligned}-g_{\mathrm{b}}(\mathbf{r}_2-\mathbf{r}_{\mathrm{b}})^2 - g_{\mathrm{d}}(\mathbf{r}_2-\mathbf{r}_{\mathrm{d}})^2 - g_{\mathrm{m}}(\mathbf{r}_2-\mathbf{r}_1)^2 = \\ -\bar{g}_{\mathrm{bdm}}(\mathbf{r}_2 - \bar{\mathbf{r}}_{\mathrm{bdm}})^2 - \frac{g_{\mathrm{bd}}{\mathbf{r}_{\mathrm{bd}}}^2 + g_{\mathrm{bm}}{\mathbf{r}_{\mathrm{bm}}}^2 + g_{\mathrm{dm}}{\mathbf{r}_{\mathrm{dm}}}^2}{\bar{g}_{\mathrm{bdm}}}.\end{aligned} \tag{55}$$

The above abbreviations are defined as:

$$\begin{aligned} g_{\mathrm{ij}} &:= g_{\mathrm{i}}g_{\mathrm{j}}; \qquad r_{\mathrm{ij}} := \mathbf{r}_{\mathrm{i}} - \mathbf{r}_{\mathrm{j}}; \qquad i,j \in b,d,m \\ \bar{g}_{\mathrm{bdm}} &:= g_{\mathrm{b}} + g_{\mathrm{d}} + g_{\mathrm{m}} \\ \bar{r}_{\mathrm{bdm}} &:= \frac{g_{\mathrm{b}}\mathbf{r}_{\mathrm{b}} + g_{\mathrm{d}}\mathbf{r}_{\mathrm{d}} + g_{\mathrm{m}}\mathbf{r}_2}{\bar{g}_{\mathrm{bdm}}},\end{aligned} \tag{56}$$

and for ease of notation we have used $\mathbf{r}_{\rm m} := \mathbf{r}_1$. The integration (54) over the simple Gaussian indicated in Eq. (55) then yields immediately

$$A_{\rm b}A_{\rm d}\left(\frac{\pi}{\bar{g}_{\rm bdm}}\right)^{\frac{3}{2}} e^{-\frac{1}{\bar{g}_{\rm bdm}}\left(g_{\rm bd}{\mathbf{r}_{\rm bd}}^2+g_{\rm bm}{\mathbf{r}_{\rm bm}}^2+g_{\rm dm}{\mathbf{r}_{\rm dm}}^2\right)}, \tag{57}$$

where the definitions

$$\mathbf{r}_{\rm bm} := \mathbf{r}_{\rm b} - \mathbf{r}_1, \qquad \mathbf{r}_{\rm dm} := \mathbf{r}_{\rm d} - \mathbf{r}_1 \tag{58}$$

make the result a Gaussian with respect to $\mathbf{r}_1$. This becomes clear in rewriting expression (57) to finally obtain for the integral (54):

$$\begin{aligned} A_{\rm b}A_{\rm d}\int d^3\mathbf{r}_2 e^{-g_{\rm b}(\mathbf{r}_2-\mathbf{r}_{\rm b})^2} e^{-g_{\rm d}(\mathbf{r}_2-\mathbf{r}_{\rm d})^2} e^{-g_{\rm m}(\mathbf{r}_2-\mathbf{r}_1)^2} &= \\ A_{\rm b}A_{\rm d}\left(\frac{\pi}{\bar{g}_{\rm bdm}}\right)^{\frac{3}{2}} e^{-\frac{g_{\rm b}g_{\rm d}}{g_{\rm b}+g_{\rm d}}{\mathbf{r}_{\rm bd}}^2} e^{-\frac{g_{\rm m}(g_{\rm b}+g_{\rm d})}{\bar{g}_{\rm bdm}}(\mathbf{r}_1-\mathbf{\Gamma}^{\rm bd})^2} \end{aligned} \tag{59}$$

with the additional shorthand

$$\mathbf{\Gamma}^{\rm bd} := \frac{g_{\rm b}\mathbf{r}_{\rm b}+g_{\rm d}\mathbf{r}_{\rm d}}{g_{\rm b}+g_{\rm d}}. \tag{60}$$

Upon insertion in the original integral over $\mathbf{r}_1$ in Eq. (33) one thus obtains another Gaussian integral with three Gaussian lobe type functions and can thus again imply in principle a formula of the type (59). Specifically Eq. (33) now reads

$$\begin{aligned} V_{\rm abcd} &= A_{\rm a}A_{\rm b}A_{\rm c}A_{\rm d}h\sum_{\rm m} f_{\rm m}\left(\frac{\pi}{\bar{g}_{\rm bdm}}\right)^{\frac{3}{2}} e^{-\frac{g_{\rm b}g_{\rm d}}{g_{\rm b}+g_{\rm d}}{\mathbf{r}_{\rm bd}}^2} \\ &\int d^3\mathbf{r}_1 e^{-g_{\rm a}(\mathbf{r}_1-\mathbf{R}_{\rm a})^2} e^{-g_{\rm c}(\mathbf{r}_1-\mathbf{R}_{\rm c})^2} e^{-\frac{g_{\rm m}(g_{\rm b}+g_{\rm d})}{\bar{g}_{\rm bdm}}(\mathbf{r}_1-\mathbf{\Gamma}^{\rm bd})^2}. \end{aligned} \tag{61}$$

Applying again formula (59) for the integral over the three Gaussian lobes with respect to $\mathbf{r}_1$ in the above expression yields (in an obvious notation following the solution of the first integral with respect to $\mathbf{r}_2$):

$$\begin{aligned} V_{\rm abcd} &= A_{\rm a}A_{\rm b}A_{\rm c}A_{\rm d}h\sum_{\rm m} f_{\rm m}\left(\frac{\pi}{\bar{g}_{\rm bdm}}\right)^{\frac{3}{2}}\left(\frac{\pi}{\bar{g}_{\rm acm}}\right)^{\frac{3}{2}} e^{-\frac{g_{\rm b}g_{\rm d}}{g_{\rm b}+g_{\rm d}}{\mathbf{r}_{\rm bd}}^2} \\ & e^{-\frac{1}{\bar{g}_{\rm acm}}\left(g_{\rm ac}{\mathbf{r}_{\rm ac}}^2+g_{\rm am}{\mathbf{r}_{\rm am}}^2+g_{\rm cm}{\mathbf{r}_{\rm cm}}^2\right)}, \end{aligned} \tag{62}$$

where the abbreviations

$$\bar{g}_{\rm acm} := g_{\rm a}+g_{\rm c}+\frac{g_{\rm m}(g_{\rm b}+g_{\rm d})}{\bar{g}_{\rm bdm}} \tag{63}$$

$$
\begin{aligned}
g_{\mathrm{am}} &:= g_{\mathrm{a}} g_{\mathrm{m}} \frac{(g_{\mathrm{b}} + g_{\mathrm{m}})}{\bar{g}_{\mathrm{bdm}}} \\
g_{\mathrm{cm}} &:= g_{\mathrm{c}} g_{\mathrm{m}} \frac{(g_{\mathrm{b}} + g_{\mathrm{m}})}{\bar{g}_{\mathrm{bdm}}} \\
\mathbf{r}_{\mathrm{am}} &:= \mathbf{R}_{\mathrm{a}} - \mathbf{\Gamma}^{\mathrm{bd}} \\
\mathbf{r}_{\mathrm{cm}} &:= \mathbf{R}_{\mathrm{c}} - \mathbf{\Gamma}^{\mathrm{bd}}
\end{aligned}
$$

were used. This completes the proof of Eq. (34,35) except that the present expression appears asymmetric in the indices of the for Gaussian lobe basis functions indexed a,b,c,d. On the other hand in Eq. (34,35) the identical, yet symmetric, formula was used. To be brief we here constrain the considerations to two examples, the other terms in Eq. (34,35) can be derived in an analogous way. We first show that

$$
\bar{g}_{\mathrm{acm}} \bar{g}_{\mathrm{bdm}} = g_0. \tag{64}
$$

In fact we have

$$
\begin{aligned}
\bar{g}_{\mathrm{acm}} \bar{g}_{\mathrm{bdm}} &= (g_{\mathrm{a}} + g_{\mathrm{c}} + \frac{g_{\mathrm{m}}(g_{\mathrm{b}} + g_{\mathrm{d}})}{\bar{g}_{\mathrm{bdm}}})(g_{\mathrm{b}} + g_{\mathrm{d}} + g_{\mathrm{m}}) \\
&= (g_{\mathrm{b}} + g_{\mathrm{d}} + g_{\mathrm{m}})(g_{\mathrm{a}} + g_{\mathrm{c}}) + g_{\mathrm{m}}(g_{\mathrm{b}} + g_{\mathrm{d}}) \\
&= (g_{\mathrm{b}} + g_{\mathrm{d}})(g_{\mathrm{a}} + g_{\mathrm{c}}) + g_{\mathrm{m}}(g_{\mathrm{b}} + g_{\mathrm{d}} + g_{\mathrm{a}} + g_{\mathrm{c}}) \\
&= g_0.
\end{aligned} \tag{65}
$$

The definitions (56,63) were used and the result is indeed $g_0$ defined in (35). As a second example we check the coefficient of ${\mathbf{R}_{\mathrm{a}}}^2$, which is found according to Eq. (62) from

$$
\frac{1}{\bar{g}_{\mathrm{acm}}}(g_{\mathrm{ac}} {\mathbf{r}_{\mathrm{ac}}}^2 + g_{\mathrm{am}} {\mathbf{r}_{\mathrm{am}}}^2) = \tag{66}
$$
$$
\frac{1}{g_{\mathrm{a}} + g_{\mathrm{c}} + g_{\mathrm{m}}}(g_{\mathrm{a}} g_{\mathrm{c}} [\mathbf{R}_{\mathrm{a}} - \mathbf{R}_{\mathrm{c}}]^2 + g_{\mathrm{a}} g_{\mathrm{m}} [\mathbf{R}_{\mathrm{a}} - \mathbf{\Gamma}^{\mathrm{bd}}]^2),
$$

so that the coefficient of ${\mathbf{R}_{\mathrm{a}}}^2$ becomes

$$
\frac{g_{\mathrm{a}} g_{\mathrm{c}} + g_{\mathrm{a}} g_{\mathrm{m}}}{\bar{g}_{\mathrm{acm}}} = \tag{67}
$$
$$
\frac{(g_{\mathrm{a}} g_{\mathrm{c}} + g_{\mathrm{a}} g_{\mathrm{m}}) \bar{g}_{\mathrm{bdm}}}{\bar{g}_{\mathrm{acm}} \bar{g}_{\mathrm{bdm}}}.
$$

The denominator has just been shown to amount to the required $g_0$ in the following of Eq. (64), and the numerator equals

$$
\left( g_{\mathrm{a}} g_{\mathrm{c}} + g_{\mathrm{a}} \frac{g_{\mathrm{m}}(g_{\mathrm{b}} + g_{\mathrm{d}})}{\bar{g}_{\mathrm{bdm}}} \right) \bar{g}_{\mathrm{bdm}} = \tag{68}
$$

$$
\begin{aligned}
g_{\rm a}g_{\rm c}(g_{\rm b}+g_{\rm d}+g_{\rm m})+g_{\rm a}g_{\rm m}(g_{\rm b}+g_{\rm d}) &= \\
g_{\rm a}\left(g_{\rm c}(g_{\rm b}+g_{\rm d})+g_{\rm m}(g_{\rm c}+g_{\rm b}+g_{\rm d})\right) &= \\
D^{\rm c;bd}&,
\end{aligned}
$$

and this establishes the correspondence to Eq. (34,35) as far as ${\bf R_a}^2$ is concerned. The other terms in the exponent of Eq. (34,35) can be traced in analogous ways which we will not further pursue here.

## VII. APPENDIX B: PARAMETERS OF EQ. (38)

Combining Eq. (34,35) with a corresponding expression for $V_{\rm a'b'c'd'}$ yields a simple Gaussian with respect to the lattice vector $\mathbf{R}$ as symbolized in Eq. (38). The parameters J, G and $\mathbf{\Gamma}$ have the form:

$$
\begin{aligned}
J &:= A_{\rm a}A_{\rm b}A_{\rm c}A_{\rm d}A_{\rm a'}A_{\rm b'}A_{\rm c'}A_{\rm d'}\frac{\pi^3}{\bar{g}_{\rm acm}^{\frac{3}{2}}}\frac{\pi^3}{\bar{g}_{\rm bdm}^{\frac{3}{2}}}e^{-({\rm E_0+E_0'})}e^{\frac{({\bf E}_1+{\bf E}_1')^2}{4(E_2+E_2')}} \qquad (69)\\
G &:= E_2+E_2'\\
\mathbf{\Gamma} &:= \frac{1}{2}\frac{{\bf E}_1+{\bf E}_1'}{E_2+E_2'}.
\end{aligned}
$$

The abbreviations are defined as:

$$
\begin{aligned}
E_0 &:= D^{\rm c;bd}g_{\rm a}{\bf R_a}^2+D^{\rm a;bd}g_{\rm c}{\bf R_c}^2+D^{\rm d;ac}g_{\rm b}{\bf R_b}^2+D^{\rm b;ac}g_{\rm d}{\bf R_d}^2 \qquad (70)\\
&\quad -2g_{\rm a}g_{\rm c}(g_{\rm b}+g_{\rm d}+g_{\rm m}){\bf R_c}{\bf R_a}-2g_{\rm b}g_{\rm d}(g_{\rm a}+g_{\rm c}+g_{\rm m}){\bf R_b}{\bf R_d}\\
&\quad -2g_{\rm m}\left(g_{\rm a}g_{\rm b}{\bf R_a}{\bf R_b}+g_{\rm a}g_{\rm d}{\bf R_a}{\bf R_d}+g_{\rm c}g_{\rm b}{\bf R_c}{\bf R_b}+g_{\rm c}g_{\rm d}{\bf R_c}{\bf R_d}\right)\\
{\bf E}_1 &:= 2g_{\rm b}\left(D^{\rm d;ac}-g_{\rm d}(g_{\rm a}+g_{\rm c}+g_{\rm m})\right){\bf R_b}+ \qquad (71)\\
&\quad 2g_{\rm d}\left(D^{\rm b;ac}-g_{\rm b}(g_{\rm a}+g_{\rm c}+g_{\rm m})\right){\bf R_d}\\
&\quad -2g_{\rm m}g_{\rm a}(g_{\rm b}+g_{\rm d}){\bf R_a}-2g_{\rm m}g_{\rm c}(g_{\rm b}+g_{\rm d}){\bf R_c}\\
E_2 &:= D^{\rm d;ac}g_{\rm b}+D^{\rm b;ac}g_{\rm d}-2g_{\rm b}g_{\rm d}(g_{\rm a}+g_{\rm c}+g_{\rm m}). \qquad (72)
\end{aligned}
$$

Analogous definitions hold for the primed quantities referring to the Gaussian with primed indices a',b',c' and d'.

---

[1] P. HOHENBERG, W. KOHN, Phys. Rev. 136: B864 (1964); W. KOHN and L. J. SHAM, Phys. Rev. 141: A1133 (1965).

[2] T. GRABO, T. KREIBICH, S. KURTH, and E. K. U. GROSS Strong Coulomb Correlations in Electronic Structure: Beyond the Local Density Approximation, edited by V. I. Anisimov (Gordon & Breach) (1998).

[3] M. PETERSILKA, U. J. GOSSMANN, and E. K. U. GROSS , Phys. Rev. Lett. **76**, 1212 (1995).

[4] Z. WU, R. E. COHEN, D. J. SINGH, R. GUPTA, and M. GUPTA, Phys. Rev. B, **61**, 16491 (2000).

[5] N. A. BESLEY and P. M. W. GILL, J. Chem. Phys. **120**, 7290 (2004).

[6] S. GOEDECKER and C. J. UMRIGAR, Phys. Rev. Lett. **81**, 886 (1998); K. YASUDA, Phys. Rev. Lett. **88**, 053001 (2002).

[7] A. GEORGES, G. KOTLIAR, W. KRAUTH, and M. J. ROZENBERG , Rev. Mod. Phys. **68**, 13 (1996).

[8] cf. e. g. R. HOTT, Phys. Rev. B **44**, 1057 (1991).

[9] G. STOLLHOFF and P. FULDE, Z. Phys. B **29**, 231 (1978).

[10] S. HORSCH, P. HORSCH, and P. FULDE, Phys. Rev. B **29**, 1870 (1984).

[11] H. STOLL, Chem. Phys. Lett. **191**, 548 (1992); H. STOLL Phys. Rev. B **46**,6700 (1992).

[12] B. PAULUS, P. FULDE, and H. STOLL, Phys. Rev. B **54**, 2556 (1996); S. KALVODA, B. PAULUS, P. FULDE, and H. STOLL, Phys. Rev. B **55**, 4027 (1997); B. PAULUS, F.-J. SHI, and H. STOLL, J. Phys. Cond. Matter **9**, 2745 (1997).

[13] K. DOLL, M. DOLG, P. FULDE, and H. STOLL, Phys. Rev. B **52**, 4842 (1995); K. DOLL, M. DOLG, and H. STOLL, Phys. Rev. B **54**, 13529 (1996); K. DOLL, M. DOLG, P. FULDE, and H. STOLL, Phys. Rev. B **55**, 10282 (1997).

[14] A. ABDURAHMAN, A. SHUKLA, and M. DOLG, Phys. Rev. B **65**, 115106 (2001).

[15] M. ALBRECHT, B. PAULUS, and H. STOLL, Phys. Rev. B **56**, 7339 (1997).

[16] J. GRÄFENSTEIN, H. STOLL, and P. FULDE, Chem. Phys. Lett. **215**, 610 (1993).

[17] J. GRÄFENSTEIN, H. STOLL, and P. FULDE, Phys. Rev. B **55**, 13588 (1997).

[18] M. ALBRECHT, P. FULDE, and H. STOLL, Chem. Phys. Lett. **319**, 355 (2000).

[19] M. ALBRECHT, P. REINHARDT, and J.–P. MALRIEU, Theor. Chem. Acc. **100**, 241 (1998).

[20] C. WILLNAUER, and U. BIRKENHEUER, J. Chem. Phys. **120**, 11910 (2004).

[21] A. ABDURAHMAN, M. ALBRECHT, A. SHUKLA, and M. DOLG, J. Chem. Phys. **110**, 8819 (1998).

[22] W. M. C. FOULKES, L. MITAS, R. J. NEEDS, and G. RAJAGOPAL, Rev. Mod. Phys. **73**, 33 (2001).

[23] A. J. WILLIAMSON, R. Q. HOOD, R. J. NEEDS, and G. RAJAGOPAL, Phys. Rev. B **57**, 12140 (1998).

[24] L. MOTAS, Comput. Phys. Commun. **96**, 107 (1996).

[25] M. D. TOWLER, R. Q. HOOD, and R. J. NEEDS, Phys. Rev. B **62**, 2330 (2000).

[26] J. SCHIRMER and F. MERTINS , Int. J. Quantum Chem. **58** (1996) 329.

[27] J. IGARASHI, P. UNGER, K. HIRAI, and P. FULDE, Phys. Rev. B **49**, 16181 (1994).

[28] M. TAKAHASHI and J. IGARASHI, Phys. Rev. B **54**, 13566 (1996).

[29] M. ALBRECHT and J. IGARASHI, J. Phys. Soc. Jap. **70**, 1035 (2001).

[30] M. ALBRECHT, Theo. Chem. Acc. **107**, 71 (2002).

[31] M. ALBRECHT and P. FULDE, Phys. Stat. Sol. b **234**, 313 (2002).

[32] M. ALBRECHT and A. SCHNURPFEIL, Phs. Stat. Sol. b **241**, 2179 (2004).

[33] M. ALBRECHT , *Towards a frequency independent incremental ab initio scheme for the self energy*, *accepted by Theor. Chem. Acc.* (2005); cond.math: 0408325.

[34] P. FULDE, *Electron Correlations in Molecules and Solids*, Springer Series in Solid-State Sciences, vol. 100, Springer, Berlin (1995).

[35] as based on *e. g.*: M. GELL–MANN, K. BRUECKNER, Phys. Rev. **106**, 364 (1957).

[36] M. HÄSER and J. ALMLÖF, J. Chem. Phys. **96**, 2179 (1992).

[37] F. MERTINS, Ann. Phys. **8**, 261 (1999).

[38] P. P. EWALD, Ann. Phys. **64**, 253 (1921).

[39] A. SHUKLA, M. DOLG, H. STOLL, and P. FULDE, Chem. Phys. Lett. **262**, 213 (1996).

[40] W. HACKBUSCH and B. N. KHOROMSKIJ , *Kronecker Tensor–Product Approximation to Certain Matrix-Valued Functions in Higher Dimensions*, *submitted* (2004).

[41] I. S. GRADSTEYN and I. M. RYZHIK, *Table of Integrals, Series and Products*, Academic Press – New York (1965); cf. formula 8.180-4.

[42] cf. formula 9.522-1 in Ref.[41].

[43] J. L. ANDERSON, J. NASISE, K. PHILIPSON, and F. E. PRETZEL, J. Phys. Chem. Solids **31**, 613 (1970).

[44] R. T. POOLE, J. G. JENKIN, J. LIESEGANG, and R. C. G. LECKEY, Phys. Rev. B **11**, 5179 (1975).

[45] H. TATEWAKI,Tatewaki, Phys. Rev. B **60**, 3777 (1999).

[46] E. L. SHIRLEY, Phys. Rev. B **58**, 9579 (1998).

[47] M. PRENCIPE, A. ZUPAN, R. DOVESI, and E. APRÁ, Phys. Rev. B **51**, 3391 (1995).

[48] C. FUTH, V. BEZUGLY, U. BIRKENHEUER, and M. ALBRECHT, *Obtaining the bandstructure of* $[(HF)_2]_\infty$ *from an incremental self energy*, *in preparation* (2005).